\documentclass[aps,prd,10pt,twocolumn,showpacs,superscriptaddress]{revtex4-1}

\usepackage{amsmath,amssymb,amsfonts}
\usepackage{slashed}

\usepackage{graphicx}
\usepackage{color}
\usepackage{hyperref}
\usepackage{dsfont}
\usepackage{subfigure}

\definecolor{darkred}{rgb}{0.8,0.1,0.1}
\hypersetup{
    colorlinks=true,
    linkcolor=darkred, 
    citecolor=blue, 
}

\newcommand{\sbr}[1]{{\scriptscriptstyle (#1)}}

\begin{document}

\title{Saturating the unitarity bound in AdS/\texorpdfstring{CFT$^{}_\text{(AdS)}$}{CFT\_(AdS)}}

\author{Thorsten Ohl}
\email{ohl@physik.uni-wuerzburg.de}
\author{Christoph F.~Uhlemann}
\email{uhlemann@physik.uni-wuerzburg.de}
\affiliation{Institut f\"ur Theoretische Physik und Astrophysik\\
  Universit\"at W\"urzburg, Am Hubland, 97074 W\"urzburg, Germany}

\date{\today}



\begin{abstract}
We investigate the holographic description of CFTs defined on the cylinder and on AdS,
which include an operator saturating the unitarity bound.
The standard Klein-Gordon field with the corresponding mass and boundary conditions
on global AdS$_{d+1}$ and on an AdS$_{d+1}$ geometry with AdS$_d$ conformal boundary
contains ghosts.
We identify a limit in which the singleton field theory is obtained 
from the bulk theory with standard renormalized inner product,
showing that a unitary bulk theory corresponding to an operator 
which saturates the unitarity bound can be formulated and that this 
yields a free field on the boundary.
The normalizability issues found for the standard Klein-Gordon field
on the geometry with AdS$_d$ conformal boundary are avoided for the
singleton theory,
which offers interesting prospects for multi-layered AdS/CFT.
\end{abstract}


\maketitle
\section{Introduction}
The AdS/CFT correspondence \cite{Maldacena:1997re, Witten:1998qj, Gubser:1998bc} 
relates a $(d\,{+}\,1)$-dimensional gravity theory with asymptotically anti-de Sitter (AdS) boundary 
conditions to a $d$-dimensional conformal field theory (CFT) on the boundary.
Besides the common example involving global AdS$_{d+1}$ and a CFT on the cylinder,
the case where the boundary theory is itself defined on AdS$_d$ is of particular interest.
It not only provides a window to strongly-coupled curved-space quantum field theory, 
but also offers the possibility of multi-layered AdS/CFT dualities.
String-theory configurations with branes ending on branes providing a dual description of CFTs on AdS 
have been discussed in \cite{Aharony:2011yc}.
The reflection of CFT unitarity properties in the dual bulk description has been investigated for 
global AdS and CFTs on the cylinder in \cite{Andrade:2011dg}, and for an AdS$_{d+1}$ geometry with AdS$_d$ conformal
boundary, where the CFT is defined, in \cite{Andrade:2011nh}.
As found there, for a Klein-Gordon field with mass and boundary conditions such that the dual operator violates the
unitarity bound \cite{UBMack,Dobrev:1985qv, Minwalla:1997ka}, the bulk theory contains ghosts.
The case where the dual operator saturates the unitarity bound is of particular interest, as fields with
that mass frequently appear in supergravity spectra on geometries relevant for AdS/CFT, see e.g.\ \cite{Kim:1985ez}.
As found in \cite{Andrade:2011dg,Andrade:2011nh} the standard Klein-Gordon field contains ghosts in that case,
although a unitary representation of the conformal group is expected to exist.
In this work we are interested in that particular case, for which the singleton field theory turns out to play a crucial role.

The singleton \cite{Dirac:1963ta,Flato:1980we,Starinets:1998dt,Angelopoulos:1999bz,Bekaert:2011js} is a particular representation of the 
isometry group SO($2$,$d$) of AdS$_{d+1}$. 
The maximal compact subgroup is SO$(2)\otimes$SO$(d)$ and representations $D(E,j)$ 
are characterized by an energy $E$ (the lowest weight of SO(2)) and a set of SO($d$) quantum numbers $j$.
The scalar singleton is realized as an indecomposable representation 
$D(d/2-1,0)\rightarrow D(d/2+1,0)$ for $d\,{>}\,2$ (see \cite{Flato:1990eu} for AdS$_3$) and 
the structure can be extended to a Gupta-Bleuler triple of scalar $\rightarrow$ physical $\rightarrow$ gauge modes 
as \cite{Flato:1980we,Araki:1985pf,Flato:1986uh}
\begin{equation}\label{eqn:GB-triple}
   D(d/2+1,0)\rightarrow D(d/2-1,0)\rightarrow D(d/2+1,0)~.
\end{equation}
Among the remarkable properties of this representation is that it allows for the construction of a gauge theory for a scalar field
with mass $m^2=-d^2/4+\nu^2$ if $\nu^2=1$.
Its role in AdS/CFT has been emphasized and discussed in \cite{Gunaydin:1998km,Duff:1998hj}.

The particular inner product used for the singleton field theory was obtained in \cite{Flato:1980we} as 
the limit $\nu\rightarrow 1$ of the non-renormalized inner product of solutions to the Klein-Gordon equation
with generic $\nu<1$.
Taking into account the contribution of the holographic counterterms to the inner product \cite{Compere:2008us} we 
identify an alternative limit yielding the singleton theory for fixed $\nu=1$.
This allows for a direct application of the standard AdS/CFT dictionary, showing that the unitary singleton describes 
a free field on the boundary
-- as expected for a field saturating the CFT unitarity bound.

We then turn to the holographic description of CFTs which are itself defined on AdS$_d$.
A geometry for a holographic description of such CFTs 
is constructed as a quotient of AdS$_{d+1}$ sliced by AdS$_d$ hypersurfaces, such that 
the conformal boundary is a single copy of AdS$_d$ \cite{Aharony:2010ay}.
There is an additional subtlety if Neumann boundary conditions
are chosen on the boundary of the AdS$_d$ hypersurfaces \cite{Andrade:2011nh}, resulting in a breaking of the bulk isometries or unitarity.
We discuss the singleton on the AdS$_d$ slicing of AdS$_{d+1}$, yielding also for that case a unitary bulk theory for $\nu=1$.
Furthermore, we find that the normalizability issues for Neumann boundary conditions on the boundary of AdS$_d$ are avoided.
Implications for multi-layered AdS/CFT dualities are discussed briefly.

The paper is organized as follows. In Sec.~\ref{sec:singleton-global} we derive the singleton theory on global AdS$_{d+1}$
from the Klein-Gordon field with renormalized inner product and discuss its role for the unitarity bound.
In Sec.~\ref{sec:singleton-adsd} we perform the same construction on the geometry with AdS$_d$
conformal boundary and discuss the normalizability issues found previously for the standard Klein-Gordon field.
We conclude and comment on prospects for multi-layered AdS/CFT in Sec.~\ref{sec:conclusions}.

\section{The singleton on \texorpdfstring{$\text{AdS}_\text{d+1}$}{AdS\_d+1} in global coordinates}\label{sec:singleton-global}
To fix notation we recapitulate in this section the standard construction of the singleton on AdS in global coordinates \cite{Flato:1980we,Starinets:1998dt}.
We also offer a new perspective on the choice of the inner product in the light of \cite{Compere:2008us}.
We choose global coordinates $(z,\tau,\Omega_{d-1})$ on AdS$_{d+1}$ such that the line element reads
\begin{align}
 ds^2=\frac{l^2}{\sin^2\!z}\big(-d\tau^2+dz^2+\cos^2\!z\, d\Omega_{d-1}^2\big)~,
\end{align}
and consider a Klein-Gordon field with mass $m^2l^2=-d^2/4+\nu^2$ and action
\begin{equation}\label{eqn:KGaction}
 S=-\frac{1}{2}\int d^{d+1}x\, \sqrt{g} \Big(g^{MN}\partial_M\phi \partial_N\phi+m^2\phi^2\Big)~.
\end{equation}
Our focus here is on the case $\nu=1$.
The standard inner product associated to (\ref{eqn:KGaction}) reads
\begin{align}\label{eqn:KG-inner-product}
 \langle \phi^{}_1,\phi^{}_2\rangle &= \int_{\Sigma}\sqrt{g_\mathrm{ind}}n^\mu\big(\phi_1^\star\stackrel{\leftrightarrow}{\partial}_\mu\phi^{}_2\big)~,
\end{align}
where $\Sigma$ is a spacelike hypersurface with unit normal vector field $n^\mu\partial_\mu$.
To solve the field equations we employ the ansatz
\begin{align}
\phi(z,\tau,\Omega_{d-1})=e^{-i \omega \tau}\, Y_{\vec{L}}(\Omega_{d-1})\, f(z)~,
\end{align}
where $Y_{\vec{L}}$ are the spherical harmonics on $S^{d-1}$ satisfying $\bigtriangleup_{S^{d-1}}Y_{\vec{L}}=-L(L+d-2)Y_{\vec{L}}$.
The resulting equation for the radial modes $f(z)$ can be written as Sturm-Liouville problem
\begin{align}\label{eqn:radial-eq}
 K f=\omega^2f~,\qquad K=\frac{1}{w(z)}\left[-\partial_zp(z)\partial_z+q(z)\right]~,
\end{align}
where $w(z)=p(z)=\cot^{d-1}\!z$ and $q(z)=\tan^{d-1}\!z \left[L(L+d-2)\cos^{-2}z+m^2l^2\sin^{-2}\!z\right]$.
Choosing for $\Sigma$ a surface of constant $\tau$ 
we find
\begin{align}\label{eqn:KG-inner-product2}
 \langle \phi^{}_1,\phi^{}_2\rangle &=\delta_{\vec{L}_1,\vec{L}_2}(\omega_1+\omega_2)e^{i(\omega_1-\omega_2)\tau}l^{d-1}\langle f^{}_1,f^{}_2\rangle_\mathrm{SL}~,
\end{align}
where the Sturm-Liouville inner product is given by
\begin{align}
 \langle f^{}_1,f^{}_2\rangle^{}_\mathrm{SL} &=\int dz\cot^{d-1}\!z\,f_{1}^\star f^{}_2~.
\end{align}
Using partial integration and (\ref{eqn:radial-eq}) yields
\begin{equation}\label{eqn:reducedSL}
 \langle f_1,f_2\rangle^{}_\mathrm{SL}=\frac{1}{\omega^{\ast 2}_1-\omega_2^2}\left[\cot^{d-1}\!z\,\big(f_1^\ast f_2^\prime-{f_1^\prime}^\ast f_2\big)\right]_0^{\pi/2}~,
\end{equation}
which is to be understood in the distributional sense.
The two independent sets of solutions to (\ref{eqn:radial-eq}) are
\begin{align}\label{eqn:SL-solutions}
 f=&\sin ^{\frac{d}{2}-\nu}\!z \cos ^L\!z\, _2F_1\Big(\frac{a- \omega -\nu}{2} ,\frac{a+\omega -\nu}{2} ; a  ; \cos ^2\!z\Big)~,
\end{align}
where $a=d/2+L$, and a second set which is $f\big\vert_{L\rightarrow 2-d-L}$ for odd $d$ and 
a combination involving explicit logarithms for even $d$ \cite{abramowitz+stegun}.
Demanding the solutions to be regular at the origin $z=\pi/2$ selects the modes (\ref{eqn:SL-solutions}).
For the singleton theory, instead of deriving the frequency spectrum from a vanishing-flux boundary condition, one imposes \cite{Starinets:1998dt}
\begin{align}\label{eqn:omega}
 \pm\omega=a-1+2n~, \qquad n\in\mathbb N\cup\lbrace 0\rbrace~.
\end{align}
The $n\geq 1$ modes are the standard Dirichlet solutions, i.e.\ they are $\mathcal O(z^{d/2+1})$ in the boundary limit.
They form the representation $D(d/2+1)$ with lowest-weight state given by the $n=1$ mode with $\omega=d/2+1$.
Adding the $n=0$ solutions yields the representation $D(d/2-1)$.
The $n=0$ mode is $\mathcal O(z^{d/2-1})$ and is not normalizable with respect to (\ref{eqn:KG-inner-product}).
Following \cite{Flato:1980we,Starinets:1998dt} we replace the radial part of the inner product by
\begin{align}\label{eqn:FF-inner-product}
 \langle f^{}_1,f^{}_2\rangle^{}_\mathrm{sing} &=\lim_{\nu\rightarrow 1_{}^-}(1-\nu)\, \langle f^{}_1,f^{}_2\rangle^{}_\mathrm{SL}~.
\end{align}
where the $f_i$ on the right hand side are the modes $f$ of (\ref{eqn:SL-solutions}) for generic $\nu<1$.
Evaluating this inner product yields 
\begin{align}\label{eqn:FF-inner-product-eval}
\langle f^{}_1,f^{}_2\rangle^{}_\mathrm{sing}=\frac{1}{2}\delta^{}_{n_1,0}\delta^{}_{n_2,0}~, 
\end{align}
i.e.\ except for the $n=0$ mode which is of positive norm, all other modes are of norm zero, i.e.\ pure gauge.
The singleton representation is induced on the quotient space 
obtained by identifying in the space spanned by the $n\geq 0$ solutions 
those which differ only by $n\geq 1$ modes.
It has only a single $(E,j)$ trajectory, hence the name.

\subsection{Relation to the Renormalized Inner Product}\label{sec:singleton-kappa-limit}
As discussed in detail in \cite{Compere:2008us} the contribution of the holographic counterterms is crucial 
for dealing with the divergences in the symplectic structure and inner product. 
We now discuss the singleton representation from that perspective.
The action (\ref{eqn:KGaction}) reduces on shell to a boundary term 
$S_\mathrm{on-shell}=\frac{1}{2}\int_{z=0}\phi\sqrt{g^{zz}}\partial_z\phi$ which is divergent for $\nu\geq 1$.
We suppress the standard volume form constructed from the (induced) metric from here on.
For $\nu=1$ $S_\mathrm{on-shell}$ contains a logarithmic divergence and is
rendered finite by regularizing the geometry with a cut-off $z\geq\epsilon$ and 
adding boundary terms at $z=\epsilon$. 
The renormalized action is $S_\mathrm{ren}:=S+S_\mathrm{ct}$ with
\begin{align}\label{eqn:Sct}
 S_\mathrm{ct}=-\frac{1}{2}\int_{z=\epsilon}
 \left[  \left(\frac{d}{2}-1\right)\phi^2
         -\big(\log z+\kappa\big)\phi\,\square_{g_\mathrm{ind}}\phi\right]~.
\end{align}
Note that we have included, with an arbitrary coefficient $\kappa$, a boundary term which is compatible with all symmetries and finite for $\nu=1$.
With the asymptotic expansion of $\phi$ given by 
\begin{equation}\label{eqn:asymptSol}
 \phi = \phi_{}^\sbr{0}z^{\frac{d}{2}-1}+\phi^\sbr{1}z^{\frac{d}{2}+1}\log z+\phi^\sbr{2}z^{\frac{d}{2}+1}+\dots
\end{equation}
the variation of the renormalized action reads
\begin{align}\label{eqn:deltaS1}
 \delta S_\mathrm{ren}&=
  \text{EOM}+\int_{z=\epsilon}
        \delta\phi^\sbr{0}\big(2\phi^\sbr{2}+(1-2\kappa)\phi^\sbr{1}\big)~.
\end{align}
The boundary conditions for a stationary action are therefore
either the Dirichlet condition $\delta \phi^\sbr{0}=0$
or the Neumann condition
\begin{align} \label{eqn:NeumannBC}
  2\phi^\sbr{2}+(1-2\kappa)\phi^\sbr{1}=0~.
\end{align}
The inner product associated to the renormalized action takes a form similar to (\ref{eqn:KG-inner-product2}).
The contribution of the counterterms to the inner product can be absorbed into a renormalized Sturm-Liouville product, which then reads
\begin{align}\label{eqn:SLrenorm}
\begin{split}
 \langle f_1 , f_2 \rangle^{}_\mathrm{ren} = &\langle f_1 , f_2 \rangle^{}_\mathrm{SL}\\
 &+ (\log z+\kappa)\cot^{d-1}\!z \sin z \,f_1^\ast f^{}_2 \big|_{z \rightarrow 0}~.
\end{split}
\end{align}

We now consider a particular limit which yields 
the frequency quantization (\ref{eqn:omega})
and the inner product (\ref{eqn:FF-inner-product-eval}) such that we obtain the singleton.
To this end we rescale the field as $\phi\rightarrow \phi^\prime=\kappa^{-1/2}\phi$ and 
perform a limit $\kappa\rightarrow\infty$. 
We consider the family of theories for $\kappa\in\mathbb R^+$.
The variation of the action reads
\begin{align}\label{eqn:deltaSkappa}\begin{split}
 \delta S=&-\frac{1}{2}\int \kappa^{-1}\delta\phi^{\prime}\big(-\square+m^2\big)\phi^\prime\\
          &+    \int_{z=\epsilon}
        \delta\phi^{\prime\sbr{0}}\big(2\kappa^{-1}\phi^{\prime\sbr{2}}+(\kappa^{-1}-2)\phi^{\prime\sbr{1}}\big)~.
\end{split}\end{align}
The bulk part has to vanish for any finite $\kappa$ and so the bulk field equation also applies 
as we consider the limit $\kappa\rightarrow \infty$.
However, had we included interaction terms in (\ref{eqn:KGaction}) they would become negligible with respect to the
quadratic part. 
The field rescaling ensures that we get a finite on-shell action.
In the boundary part of the variation the $\kappa^{-1}$-terms become negligible with respect to
the remaining term, so the variation reduces to 
\begin{equation}
\delta S_\mathrm{ren}=  \text{EOM}-2\int_{z=\epsilon}
        \delta\phi^{\prime\sbr{0}}\phi^{\prime\sbr{1}}~.
\end{equation}
The 
Neumann boundary condition (\ref{eqn:NeumannBC}) thus becomes $\phi^{\prime\sbr{1}}=0$.
With the expansion $f=z^{-d/2-1}(f^\sbr{0}+f^\sbr{1}z^2\log{z}+f^\sbr{2}z^2+\dots)$ we then have to solve $f^\sbr{1}=0$.
For the modes (\ref{eqn:SL-solutions}) with $\nu=1$ we have
\begin{equation}
 f^\sbr{1}=\frac{2 \Gamma (a)}{\Gamma \left(\frac{a-\omega -1}{2} \right) \Gamma \left(\frac{a+\omega -1}{2} \right)}~,
\end{equation}
which yields the frequency quantization (\ref{eqn:omega}) upon demanding the $\Gamma$-functions in the denominator to have a pole.
Note that (\ref{eqn:omega}) can thus be understood as solving vanishing-flux boundary conditions for the renormalized symplectic structure,
see Section 2 of \cite{Compere:2008us}.
For the inner product we find 
\begin{align}
 \langle \phi^{\prime}_1,\phi^{\prime}_2\rangle = &\lim_{\kappa\rightarrow\infty} \delta_{\vec{L}_1,\vec{L}_2}(\omega_1+\omega_2)e^{i(\omega_1-\omega_2)\tau}l^{d-1}\langle f^{}_1,f^{}_2\rangle_{\kappa}~,
\end{align}
where $\langle f^{}_1,f^{}_2\rangle_{\kappa}=\kappa^{-1}\langle f^{}_1,f^{}_2\rangle_\mathrm{ren}$.
With the notation $\tilde f_n:=f\big\vert_{\omega=(d-2)/2+L+2n}$ we find for the radial part
\begin{subequations}\label{eqn:innerprod-kappa}
\begin{align}
 \langle \tilde f_0, \tilde f_0\rangle_{\kappa} &=  \frac{1}{2\kappa} \left(2\kappa-\psi ^{(0)}(a) -\gamma \right)~,\\
 \langle \tilde f_n, \tilde f_m\rangle_{\kappa} &=\delta_{nm}\frac{n (a+n-1)}{2\kappa (a+2n-1)} B(n,a)^2~,\\
 \langle \tilde f_0, \tilde f_n\rangle_{\kappa} &= -\frac{1}{2\kappa} (-1)^n B(a,n)~,
\end{align}
\end{subequations}
where $n,m>0$ and $B(x,y)$ is the Euler beta function. 
Clearly, in the limit $\kappa\rightarrow\infty$ only 
$\langle \tilde f_0, \tilde f_0\rangle_{\kappa}$ is non-vanishing and in fact positive, such that we recover (\ref{eqn:FF-inner-product-eval})
up to an overall factor.

This can also be understood from a scaling argument as follows. 
We argued above that the action does not simply reduce to the boundary terms for 
$\kappa\rightarrow\infty$, as the bulk field equation applies for any finite $\kappa$ 
while the boundary terms merely affect the boundary conditions. 
However, the inner product associated to the renormalized action is just the sum of the 
bulk part (\ref{eqn:KG-inner-product}) and the boundary contributions derived from (\ref{eqn:Sct}). 
Thus, it indeed reduces to the boundary part arising from the term proportional to $\kappa$ as we 
take the limit $\kappa\rightarrow\infty$ with the corresponding field rescaling.
This remaining part now vanishes for the $n>0$ modes as they satisfy the standard Dirichlet boundary condition.

\subsection{AdS/CFT at the unitarity bound}\label{sec:adscft-at-bound}

Realizing the singleton as discussed in the previous section allows for a direct interpretation 
in the AdS/CFT context.
Fluctuations of a scalar with Neumann boundary condition correspond to a deformation
of the dual CFT by an operator $\mathcal O$ with scaling dimension $d/2-\nu$ \cite{Witten:2001ua, Berkooz:2002ug}.
Performing the Legendre transform
\begin{align}
 S_\mathrm{ren}\rightarrow S_\mathrm{ren}^\mathrm{N}:=S_\mathrm{ren}-\int_{z=\epsilon}\phi^\sbr{0}\big(2\phi^\sbr{2}+(1-2\kappa)\phi^\sbr{1}\big)
\end{align}
we find
\begin{align}
 \delta S_\mathrm{ren}^\mathrm{N}=\text{EOM}-
\int_{z=\epsilon}
        \phi^\sbr{0}\,\delta\big(2\phi^\sbr{2}+(1-2\kappa)\phi^\sbr{1}\big)~,
\end{align}
and the on-shell action becomes a functional of the Neumann boundary data $2\phi^\sbr{2}+(1-2\kappa)\phi^\sbr{1}$. 
For $\kappa\rightarrow\infty$ with the field rescaling
$\phi\rightarrow \phi^\prime=\kappa^{-1/2}\phi$ discussed above, 
we find 
$\delta S^\mathrm{N}_\mathrm{ren}=\text{EOM}+\int_{z=\epsilon}2\phi^{\prime\sbr{0}}\delta\phi^{\prime\sbr{1}}$.
Following the familiar AdS/CFT identification of bulk partition function and the generating functional for boundary
correlation functions, functional differentiation of $S_\mathrm{ren}^\mathrm{N}$ with respect to $\phi^{\prime\sbr{1}}$ 
yields the connected correlation functions of the dual operator $\mathcal O$ of the CFT. 
We find
\begin{align}\label{eqn:correlators}
 \langle\mathcal O\rangle&=\frac{1}{\sqrt{g}}\frac{\delta S_\mathrm{ren}^\mathrm{N}}{\delta \phi^{\prime\sbr{1}}} = 2\phi^{\prime\sbr{0}}~,
\quad
 \langle\mathcal O \mathcal O\rangle=\frac{1}{\sqrt{g}}\frac{\delta \langle\mathcal O\rangle}{\delta \phi^{\prime\sbr{1}}} ~,
\end{align}
where 
$\phi^{\prime\sbr{1}}=-\frac{1}{2}\big(\square_{g^\sbr{0}}-\frac{1}{4}\frac{d-2}{d-1} R[g^\sbr{0}]\big)\phi^{\prime\sbr{0}}$
for a generic asymptotically-AdS
metric of the form $r^{-2}(dr\otimes dr-g)$.
The $n$-point functions with $n\geq 3$ vanish unless interactions of the bulk scalar are included.
However, for the singleton there are no gauge-invariant bulk interactions as the field is gauge-equivalent to zero in 
any compact region, so the higher correlation functions vanish.
This is characteristic of (generalized) free fields and
the singleton therefore yields the dual description of a free field on the boundary, consistent with the fact that this 
is the only way of realizing a unitary representation of the conformal group for $\Delta=d/2-1$.

\section{The Singleton on the \texorpdfstring{$\text{AdS}_\text{d}$}{AdS\_d} slicing of \texorpdfstring{$\text{AdS}_\text{d+1}$}{AdS\_d+1} }
\label{sec:singleton-adsd}
We now turn to the holographic description of CFTs defined on AdS$_d$.
A geometry for the dual description has been proposed in \cite{Aharony:2010ay}
and was discussed in the context of unitarity from the holographic perspective in \cite{Andrade:2011nh}.
As shown there, the standard Klein-Gordon theory yields ghosts for $\nu\geq 1$ and the 
renormalization turns out to be nontrivial if Neumann boundary conditions are chosen
at the boundary of AdS$_d$. We come back to that issue at the end of the section.

\subsection{The geometry}
\begin{figure}[htb]
\center
\subfigure[][]{ \label{fig:AdSslicing}
  \includegraphics[width=0.42\linewidth]{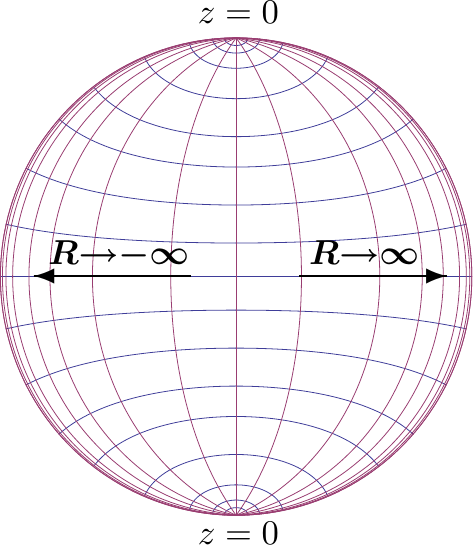}
}\qquad\qquad
\subfigure[][]{ \label{fig:adsads-bndy}
  \includegraphics[width=0.33\linewidth]{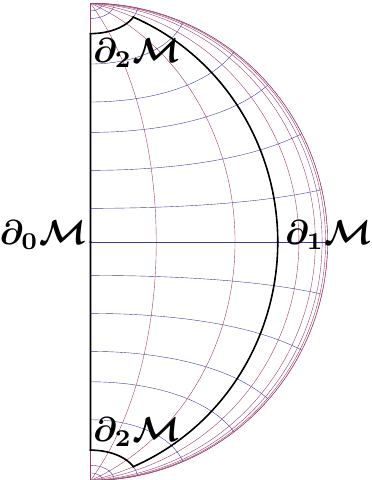}
}
\caption{The Poincar\'e disk representation of AdS$_{d+1}$ sliced by AdS$_d$ hypersurfaces is shown in Fig.~\ref{fig:AdSslicing}.
         Horizontal/vertical curves correspond to constant $z$/$R$.
         The boundary consists of two copies of AdS$_{d}$ joined at their boundaries at $z\to 0$.
         Fig.~\ref{fig:adsads-bndy} shows the boundary of the regularized geometry.
        }
\end{figure}

To obtain the slicing of AdS$_{d+1}$ with curvature radius $L$ by AdS$_d$ hypersurfaces with curvature radius $l$
we start from global coordinates in which the line element is 
\begin{equation}\label{eqn:ads-global-metric}
 ds^2=-\big(1+\rho^2/L^2\big)dt^2+\frac{1}{1+\rho^2/L^2}d\rho^2+\rho^2 d\Omega_{d-1}^2~.
\end{equation}
Parametrizing $d\Omega_{d-1}^2=d\zeta^2+\sin^2\zeta\, d\Omega_{d-2}^2$, the AdS$_d$ slicing is obtained by the coordinate transformation
$(\rho,\zeta,t)\rightarrow (R,z,\tau)$ with $t=L\tau$ and
\begin{subequations}
\begin{align}\label{eqn:coord-transf-ads}
 \frac{\rho^2}{L^2}&=\csc^2z\cosh^2 \frac{R}{L}-1~,\\
 \rho^2\sin^2\zeta&=L^2\cot^2z \cosh^2 \frac{R}{L}~.\label{eqn:coord-transf-ads2}
\end{align}
\end{subequations}
The resulting line element reads
\begin{equation}
 ds^2=dR^2+\frac{L^2}{l^2}\cosh^2\frac{R}{L}\:ds^2_\text{AdS$_d$}~,
\end{equation}
where
$ds^2_\text{AdS$_d$}=l^2\csc^2\!z\,(-d\tau^2+dz^2+\cos^2\!z\, d\Omega_{d-2}^2)$.
As the domain for the sine in (\ref{eqn:coord-transf-ads2}) is either $\zeta\in[0,\pi/2)$ or $\zeta\in(\pi/2,\pi]$,
two patches are needed to cover the full AdS$_{d+1}$.
The result is a geometry with conformal boundary consisting of two copies of AdS$_d$, see Fig.~\ref{fig:AdSslicing}.
A geometry with a single AdS$_d$ conformal boundary is obtained by taking a $\mathbb Z_2$ quotient identifying the two patches.
This implies that we have to choose a definite $\mathbb Z_2$ parity for the Klein-Gordon field, which imposes boundary
conditions at $R=0$.

As usual this geometry needs to be regularized to account for divergences in the on-shell action and inner products.
This was done in \cite{Andrade:2011nh} by imposing cut-offs on $y:=2le^{-R/L}$ and $z$, i.e.\ $y\geq \epsilon_1$, $z\geq \epsilon_2$.
The resulting geometry with its boundary is illustrated in Fig.~\ref{fig:adsads-bndy}.
The renormalized action  $S_\mathrm{ren}:=S+S_\mathrm{ct}$ was constructed in 
\cite{Andrade:2011nh} for $L=l=1$, which we fix henceforth, with the counterterms
\begin{align}
S^{}_\mathrm{ct}=&
 -\frac{1}{2}\int_{\partial^{}_1\mathcal M}
 \left[  \left(\frac{d}{2}-1\right)\phi^2
         -\big(\log y+\kappa\big)\phi\,\square^W_{g_\mathrm{ind}}\phi\right] \nonumber\\&
  -\frac{1}{2}\int_{\partial\partial\mathcal M} (\log y+\kappa)\phi\mathcal L_{n}\phi~.\label{eqn:SctdMnu1}
\end{align}
The associated inner product reads
\begin{align}
 \langle\phi_1,\phi_2\rangle^{}_\mathrm{ren}&=\langle\phi_1,\phi_2\rangle^{}_{\mathcal M}
+\big(\log\epsilon_1+\kappa\big)\langle\phi_1,\phi_2\rangle^{}_{\partial^{}_1\mathcal M}~,\label{eqn:ren-prod-nu1}
\end{align}
where $\langle\cdot,\cdot\rangle_{\partial_1\mathcal M}$ denotes the AdS$_d$ inner product evaluated at fixed
$y=\epsilon_1$.

\subsection{AdS/\texorpdfstring{CFT$_\text{AdS}$}{CFT\_AdS} at the unitarity bound -- the singleton}
We now construct the singleton theory on this geometry analogously to the construction in Section \ref{sec:singleton-kappa-limit}.
Dropping terms which vanish upon imposing the field equations or $\mathbb Z_2$ parity, 
the variation of $S_\mathrm{ren}$ reads
\begin{align}\label{eqn:deltaS1AdS}
 \delta S_\mathrm{ren}=&
               \int\limits_{\partial^{}_2\mathcal M}\!\! \delta\phi\sqrt{g^{zz}}\partial_z\phi
      +\int\limits_{\partial_1^{}\mathcal M}\!\!
        \delta\phi^\sbr{0}\big(2\phi^\sbr{2}+(1-2\kappa)\phi^\sbr{1}\big)
       \nonumber\\&+\int\limits_{\partial\partial\mathcal M}
           (\log y+\kappa)\delta\phi \sqrt{g^{zz}}\partial^{}_z \phi
              ~.
\end{align}
where $\partial\partial\mathcal M=\partial_1\mathcal M\cap\partial_2\mathcal M$ and similar to (\ref{eqn:asymptSol})
\begin{align}
  \phi = \phi_{}^\sbr{0}y^{\frac{d}{2}-1}+\phi^\sbr{1}y^{\frac{d}{2}+1}\log y+\phi^\sbr{2}y^{\frac{d}{2}+1}+\dots~.
\end{align}
Demanding the $\partial_1\mathcal M$ boundary term to vanish imposes boundary conditions on $f$ and we choose
the Neumann condition 
\begin{align}\label{eqn:NeumannAdS}
 2\phi^\sbr{2}+(1-2\kappa)\phi^\sbr{1}=0~.
\end{align}

As discussed in detail in \cite{Andrade:2011nh} the Klein-Gordon equation is conveniently solved by a separation ansatz 
$\phi=\varphi f$ where $\varphi$ satisfies an AdS$_d$ Klein-Gordon equation with mass $M^2=-(d-1)^2/4+\mu^2$.
The independent solutions to the radial part are 
\begin{align}\label{eqn:radial-modes-nu1}
  f^{}_i&=u^{2c_i-\frac{3}{2}}\left(1-u^2\right)^{\frac{d+2}{4}} {}_2F_1\big(c_i-\frac{\mu}{2},c_i+\frac{\mu}{2};2c_i-1;u^2\big)~,
\end{align}
where $i=1,2$~, $u=\tanh(R)$ and $c_1=3/4$, $c_2=5/4$.
$f_1$ and $f_2$ have even and odd $\mathbb Z_2$ parity, respectively.
Rescaling $\phi\rightarrow \phi^\prime=\kappa^{-1/2}\phi$ and considering the limit $\kappa\rightarrow\infty$
the $\partial_1\mathcal M$ term of the variation (\ref{eqn:deltaS1AdS}) becomes 
$\delta S_\mathrm{ren}\vert^{}_{\partial_1\mathcal M}=-2\int_{\partial_1^{}\mathcal M}
        \delta\phi^{\prime\sbr{0}}\phi^{\prime\sbr{1}}$
and the boundary condition (\ref{eqn:NeumannAdS}) becomes $\phi^{\prime\sbr{1}}=0$, demanding
the $\log$-term in the expansion of $f_i(y)$ around $y=0$ to vanish. 
This yields the spectrum of $\mu$ for which we find
$\mu=1/2$ and $\mu=2(c_i+n)$ with $n\in\mathds{N}\cup\lbrace 0\rbrace$.
Solutions corresponding to the latter choice of $\mu$ are subdominant in the boundary limit.

For the solutions constructed by means of our separation ansatz we find
\begin{align}\label{eqn:inner-prod-ads-fact}
 \langle\phi^\prime_1,\phi^\prime_2\rangle^{}_\mathrm{ren}&=\langle\varphi_1,\varphi_2\rangle \, \langle f_1, f_2\rangle_\kappa~,
\end{align}
where $\langle\varphi_1,\varphi_2\rangle$ is the standard AdS$_d$ inner product and
 $\langle f_1, f_2\rangle_\kappa=\kappa^{-1}\langle f_1, f_2\rangle_\mathrm{ren}$.
With $\tilde\kappa=\kappa+\log 2$ the renormalized Sturm-Liouville 
inner product with the counterterm contributions from (\ref{eqn:SctdMnu1}) is given by
\begin{align}\label{eqn:SLrenormAdS}
    \langle f_1 , f_2 \rangle^{}_\mathrm{ren} = &
    \langle f_1 , f_2 \rangle^{}_\mathrm{SL}  - (R-\tilde\kappa)\cosh^{d-2}\!R\,f_1^\ast f^{}_2 \big|_{R \rightarrow \infty}~.
\end{align}
Denoting $\tilde f_i:=f_i\vert_{\mu=1/2}$ and $\tilde f_i^n:=f_i\vert_{\mu=2(c_i+n)}$ we find 
\begin{subequations}\label{eqn:radial-prod-ads}
\begin{align}
 || \tilde f_i||_\kappa^2&=\frac{1}{\kappa}\Big(\tilde\kappa+\frac{3}{2}-2c_i\Big)~,\\
 \langle \tilde f_i,\tilde f_i^n\rangle_\kappa&= \frac{\sqrt{2\pi } (-1)^n n!}{2^{2c_i}\kappa\Gamma (2 c_i+n)}~,\\
 \langle \tilde f^n_i,\tilde f_i^m\rangle_\kappa &= \delta_{nm}\frac{2\pi (n!)^2 \left(2 c_i n+2 c_i+n^2-1\right)}{2^{4 c_i}\kappa(c_i+n) \Gamma (2 c_i+n)^2}~.
\end{align}
\end{subequations}
Clearly, for $\kappa\rightarrow\infty$ only $|| \tilde f_i||_\kappa^2$ is non-vanishing and in fact positive.
Thus, in that limit all the subdominant modes $\tilde f_i^n$ become pure gauge while the dominant $\tilde f_i$
remains physical, and we obtain the singleton field on the geometry with AdS on the boundary.
The choice of $\mathbb Z_2$ parity has little effect -- it only alters the spectrum of gauge modes and the form
of the radial profile of the physical $\mu=1/2$ mode close to the $\mathbb Z_2$-fixed hypersurface at $R=0$.

In \cite{Andrade:2011nh} it was shown that pushing the bulk scalar on the geometry considered here beyond the
unitarity bound yields ghosts in the spectrum. 
Likewise, ghosts were also found for the standard Klein-Gordon field with mass such that the dual operator
saturates the unitarity bound,
although in that case a unitary representation is expected to exist.
The discussion of the correlation functions of the dual CFT obtained from the singleton theory in
Section \ref{sec:adscft-at-bound} immediately applies to the singleton theory
on the geometry considered in this section. 
Thus, the singleton yields the unitary bulk dual of a boundary free field also for the dual
CFT defined on AdS$_d$, completing the discussion in \cite{Andrade:2011nh}.
Furthermore, it offers a way to avoid the issues with normalizability found there
for Neumann boundary conditions along $z$, as we discuss in more detail now.

\subsection{Renormalization and Neumann\texorpdfstring{$_\text{d}$}{-d} boundary conditions}
The normalizability issues found in \cite{Andrade:2011nh} for Neumann boundary conditions at $z=0$
(`Neumann$_d$')
for the standard Klein-Gordon field
arise for any choice of the bulk mass and
are rooted in the AdS$_d$ factor of the inner product
$\langle\phi_1,\phi_2\rangle=\langle\varphi_1,\varphi_2\rangle\,\langle f_1,f_2\rangle$.
Depending on the renormalization it either fails to be finite on the full solution space
or becomes indefinite for Neumann boundary conditions along $z$.
The result is either a drastic truncation of the spectrum of AdS$_d$ modes such that 
the bulk field fails to carry a representation of the AdS isometries,
or the appearance of ghosts such that it fails to carry a unitary one.
More precisely, the solutions we constructed by means of the separation ansatz $\phi=\varphi f$
comprise an infinite series of AdS$_d$ modes corresponding to $\mu=1/2$ and $\mu=2(c_i+n)$ with
the associated radial modes.
The AdS$_d$ factor $\langle\varphi_1,\varphi_2\rangle$ of the inner product is divergent for 
the $\mu^2\geq1$ solutions, leaving only a drastically reduced set of normalizable modes.
On the other hand, rendering that part of the inner product finite by adding counterterms
on $\partial_2\mathcal M$ -- if possible -- would spoil positive definiteness of the inner product.
The special structure of the singleton field theory automatically avoids these issues.
In fact, since the radial part of the norm vanishes for all $\mu^2>1$ modes, finiteness 
of the inner products as the cut-offs on $y$ and $z$ are removed does not require any 
additional counterterm contributions to the AdS$_d$ factor.
For the physical $\mu=\frac{1}{2}$ mode the AdS$_d$ factor of the norm is positive and so is the
radial part (\ref{eqn:radial-prod-ads}).
We thus have a well-defined semidefinite inner product on the set of all modes also for Neumann
boundary conditions along $z$ and the drastic reduction of the spectrum of AdS$_d$ modes 
found in \cite{Andrade:2011nh} is avoided.
Although promoting the $\mu^2>1$ modes to pure gauge in the $\kappa\rightarrow\infty$ limit is in fact 
a similar reduction of the physical spectrum, this way of realizing the Neumann boundary condition is 
compatible with the symmetries and with unitarity.

\section{Conclusions}\label{sec:conclusions}
The unitarity properties of CFTs on the cylinder and on AdS have been investigated from
the holographic perspective in \cite{Andrade:2011dg,Andrade:2011nh}.
As found there, the standard Klein-Gordon field yields ghosts in the bulk for mass and
boundary condition such that the dual operator saturates the unitarity bound,
although a unitary representation of the conformal group exists.
In this work we have obtained the singleton field theory as a particular limit of the Klein-Gordon
field with standard renormalized inner product, which allows for a direct AdS/CFT interpretation
showing that it provides the dual description of a free field saturating the unitarity bound.
This extends the thorough discussion of unitarity from the holographic
perspective for global AdS in \cite{Andrade:2011dg} to the case where the unitarity bound 
is saturated.

We then formulated the singleton field theory on the geometry with AdS$_d$ on the conformal 
boundary of AdS$_{d+1}$, extending
the discussion of unitarity in \cite{Andrade:2011nh} accordingly.
Remarkably, the singleton field on the AdS$_d$ slicing of AdS$_{d+1}$ does not suffer from
the normalizability issues found for Neumann$_d$ boundary conditions in \cite{Andrade:2011nh}.
This offers interesting prospects for multi-layered AdS/CFT dualities.
These were speculated to be possible for AdS on the boundary already in \cite{Compere:2008us}
and were also discussed in \cite{Andrade:2011nh}, where the normalizability issues for 
nested Neumann conditions were found to pose a challenge to explicit realizations.
In a very concrete form, the possibility of iterated AdS/CFT appears for 
topologically gauged ABJM theory, which is obtained by coupling  
the $N\,{=}\,6$ supersymmetric Chern-Simons theory \cite{Aharony:2008ug}, 
which is understood as worldvolume theory of M2 branes and admits a dual description in terms
of M-theory on AdS$_4\,{\times}\,$S$^7/\mathbb{Z}_k$,
to conformal supergravity \cite{Chu:2009gi}.
The resulting theory admits a  Higgsing to a theory for D2 branes which has AdS$_3$ as a vacuum 
solution \cite{Chu:2010fk}, thus offering 
the possibility of nested AdS/CFT \cite{Nilsson:2012ky}.
One may hope to obtain a chain of dualities relating M-theory
on AdS$_4$ to the Higgsed topologically gauged ABJM, which itself is a gravitational
theory on AdS$_3$ and may therefore have a CFT$_2$ dual.
Remarkably, Chern-Simons theory can be formulated as a singleton 
theory \cite{Flato:1990eu}, and an explicit CFT$_2$ description has been 
discussed in \cite{HarunarRashid:1991bv,Michea:1999va}.
The same may also be possible for gravity \cite{Flato:1998iy}.
Our discussion of the singleton on the geometry with AdS$_d$ boundary may therefore also be relevant for a 
concrete realization of multi-layered AdS/CFT.

\vfill
\begin{acknowledgments}
We thank Don Marolf and  Tom\'{a}s Andrade for useful comments.
CFU is supported by the German National Academic Foundation 
(Studienstiftung des deutschen Volkes)
and by Deutsche Forschungsgemeinschaft through the Research Training Group GRK\,1147 
\textit{Theoretical Astrophysics and Particle Physics}.
\end{acknowledgments}


\bibliography{singleton.bib}

\end{document}